\documentclass[apj]{emulateapj}
\usepackage{graphicx}
\usepackage{epstopdf}
\usepackage{amsmath}
\usepackage{amssymb}

\shorttitle{UFO driven warm absorbers}
\shortauthors{Bu et al.}
\begin{document}
\title{Can warm absorbers be driven by ultra-fast outflows?}
\author{De-Fu Bu\altaffilmark{1}, Xiao-Hong Yang\altaffilmark{2}}

\altaffiltext{1}{Key Laboratory for Research in Galaxies and Cosmology, Shanghai Astronomical Observatory, Chinese Academy of Sciences, 80 Nandan Road, Shanghai 200030, China; dfbu@shao.ac.cn (corresponding author 1: De-Fu Bu) }
\altaffiltext{2}{Department of physics, Chongqing University, Chongqing, 400044; yangxh@cqu.edu.cn (corresponding author 2: Xiao-Hong Yang)}




%
\begin{abstract}
Warm absorbers (WAs) located approximately in the region of $1-1000$ parsecs are common phenomena in many active galactic nuclei (AGNs). The driving mechanism of WAs is still under debate. Ultra-fast outflows (UFOs) which are launched very close to the central black hole are also frequently observed in AGNs. When UFOs move outwards, they will collide with the interstellar medius (ISM) gas. In this paper, we study the possibility that whether WAs can be generated by the interaction between ISM gas and the UFOs. We find that under some ISM gas conditions, WAs can be generated. However, the covering factor of WAs is much smaller than that given by observations. This indicates that other mechanisms should also be at work. We also find that the properties of the WAs mainly depend on the density of the ISM injected into the computational domain from the outer radial boundary (1000 parsec). The higher the density of the ISM is, the higher the mass flux and kinetic power of the WAs will be. The kinetic power of the UFO driven WAs is much less than $1\%$ of the bolometric luminosity of its host AGNs. Therefore, the UFO driven WAs might not contribute sufficient feedback to its host galaxy.
\end{abstract}


\section{Introduction}
Outflows are very common phenomena of AGNs. They are frequently observed through blue-shifted absorption lines both in UV and X-ray bands (see a review by Laha et al. 2021). In the UV bands, the absorption lines arise from the ions $\rm C_{ IV}$, $\rm N_{ V}$, $\rm O_{ VI}$, $\rm Si_{ IV}$, $\rm C_{ II}$, $\rm Si_{ II}$ and $\rm Fe_{II}$. In the hard X-ray bands, highly ionized outflows are observed through blue-shifted $\rm Fe _{XXV}$ and $\rm Fe _{XXVI}$ K-shell absorption lines (e.g., Pounds et al. 2003; Tombesi et al. 2010, 2011, 2012; Gofford et al. 2013, 2015). The velocity of the hard X-ray absorbers is in the range $0.03-0.3 c$ (with $c$ being speed of light). The mean value is $0.1 c$. The column density is in the range $N_{\rm H} \sim 10^{22}-10^{24} {\rm cm^{-2}}$. Due to its highly velocity, the highly ionized outflows are named ultra-fast outflows (UFOs). The UFOs are believed to originate from the black hole accretion disk. The launching location of UFOs is around hundreds of Schwarzschild radius from the central black hole. The UFOs can be magneto-centrifugally driven (Blandford \& Payne 1982; Fukumura et al. 2015, 2018a) or driven by radiation force due to spectral lines (Murray \& Chiang 1995; Proga et al. 2000; Nomura et al. 2016; Nomora \& Ohsuga 2017; Yang et al. 2021).

In the soft X-ray bands, slowly moving warm gas outflows (warm absorbers) with velocity $100-2000 {\rm km \ s^{-1}}$ are also frequently observed (e.g, Halpern 1984; Reynolds 1997; George et al. 1998; Canizares \& Kruper 1984; Madejski et al. 1991; Kaastra et al. 2000; Kaspi et al. 2000; Reeves et al. 2004; Sako et al. 2001; Behar et al. 2001; Blustin et al. 2004, 2005; McKernan et al. 2007; Laha et al. 2014, 2016). WAs are just mildly ionized with the ionization parameter $\log \xi$ (see Equation 9 for definition) in the range $-1-3 \ {\rm erg\ s^{-1}}$. The column density is in the range $N_{\rm H} \sim 10^{20}-10^{22.5} {\rm cm^{-2}}$. The observed location of WAs is $1-1000 {\rm \ parsec}$ (Blustin et al. 2005).

Two models have been proposed to explain the launching mechanism of WAs. The first one is the magneto-centrifugal driving outflow model (Blandford \& Payne 1982). Fukumura et al. (2018b) have used this magnetohydrodynamic (MHD) scenario to explain the WAs observed in NGC 3783. It is found that the WAs features can be well explained by the MHD scenario if they assume proper radial wind number density distribution and viewing angle.

The second one is the thermal driving outflow model (Begelman et al. 1983; Woods et al. 1996). The accretion disk can be heated up by the X-rays from the region very close to the black hole. When the sound speed of the gas at the disk surface exceeds the local escape velocity, the gas will form outflows. The thermal driving model can be applied to the region far from the black hole, where the escape velocity of outflows is very low. Owing to the slow velocity and low ionization level of WAs, several works have suggested a thermal origin of WAs (Krolik \& Kriss 1995; Krolik \& Kriss 2001; Dorodnitsyn et al. 2008; Mizumoto et al. 2019).

In this paper, we propose a third model for the production of WAs. The velocity of UFOs generated from the black hole accretion disk is much higher than the escape velocity at the location where it is launched. Therefore, the UFOs are capable to move a large distance to interact with the ISM. In this case, the energy/momentum of UFOs can be transferred to the ISM (Pounds \& King 2013). There is the possibility that the WAs are the ISM driven by UFOs. In this paper, we investigate under what ISM conditions the WAs can be driven by UFOs. We also compare the properties of the UFO driven WAs to observations. The purpose of the comparison is to study whether the scenario proposed in this paper is sufficient to produce the observed WAs.

We organize the paper as follows. In Section 2, we introduce our methods including the equations, numerical settings, physical assumptions and initial and boundary conditions of the simulations. In Section 3, we introduce our results. In Section 4, we give discussions. We summarize our results in section 5.

\section{Numerical method }

\subsection{Parameters of UFOs}
Observations show that the mass flux of UFOs is comparable to the accretion rate onto the black hole (Gofford et al. 2015). Therefore, in this paper, we assume that the mass flux of UFOs equals to the accretion rate onto the black hole. We set the accretion rate of the black hole by using the bolometic luminosity ($L_{\rm bol}$) of the AGNs. The black hole accretion rate is calculated by $\dot M_{\rm BH} = L_{\rm bol}/(\eta c^2)$, with $\eta=0.1$ being the radiative efficiency. In some models we set $L_{\rm bol} = 0.1 L_{\rm Edd}$, with $L_{\rm Edd}$ being Eddingtong luminosity. In other models, we set $L_{\rm bol} = 0.5 L_{\rm Edd}$ (see Table 1). One can also calculate the accretion rate of the black hole based on the accretion rate at the inner radial boundary. In such case, because our inner radial boundary is far away from the black hole, one need to assume the radial profile of the mass inflow rate in the region between our inner radial boundary and the black hole horizon. This will complicate the problem. Therefore, in this paper, for simplicity, we give a fixed value to the black hole accretion rate (or $L_{\rm bol}$) and assume that it does not vary with time.

Tombesi et al. (2011) found that the mean and maximum velocities of UFOs are $0.1c$ and $0.3c$, respectively. In this paper, we set velocities of UFOs either equal to $0.1 c$ or $0.3 c$.

The third important parameter of UFOs is its opening angle. We set the opening angle of the UFOs according to the observations of UFOs in some radio-loud AGNs. Tombesi et al. (2014) analyzed the properties of UFOs in some radio-loud AGNs. The inclination angel of the jet in these AGNs is known. Therefore, it is easy to know the angle at which UFOs are present. It is reported that UFOs are detected over a wide range of jet inclination angles ($\sim 10^\circ - 70^\circ$; see also Mehdipour \& Costantini 2019). In our models, we inject UFOs at the inner boundary in the region $10^\circ \leq \theta \leq 70^\circ$ (see the definition of $\theta$  below). The density and velocity of UFOs are assumed to be uniformly distributed in this angular region.

The sound speed of UFOs gas should be significantly smaller than its bulk velocity. Therefore, the internal energy flux of UFOs should be significantly smaller than its kinetic energy flux. In this paper, we set the temperature of the UFOs injected at the inner radial boundary equals to the gas temperature at its adjacent active zone. We have done test by setting a fixed temperature to UFOs, we find that the effects of temperature of UFOs on the results are negligibly small.

\subsection{Equations}
The mass of the central black hole is set to be $M_{\rm BH}=10^9M_{\odot}$, with $M_{\odot}$ being solar mass. We carry out two-dimensional hydrodynamic simulations by using the ZEUS-MP code (Hayes et al. 2006). The spherical coordinates ($r$,$\theta$,$\phi$) are employed. The equations solved are as follows,
\begin{equation}
 \frac{d\rho}{dt} + \rho \nabla \cdot {\bf v} = 0,
\end{equation}
\begin{equation}
 \rho \frac{d{\bf v}}{dt} = -\nabla p - \rho \nabla \Phi
\end{equation}
\begin{equation}
 \rho \frac{d(e/\rho)}{dt} = -p\nabla \cdot {\bf v} + \rho \dot E
\end{equation}
$\rho$, $\bf v$, $e$, $p$, $\Phi$, $\dot E$ are respectively gas density, velocity, gas internal energy per unit volume, gas pressure, gravitational potential of the central black hole, and the net heating/cooling rate per unit mass. Ideal gas equation of state $p=(\gamma-1)e$ is adopted, with $\gamma=5/3$ .

For an AGN, it is generally believed that a compact hot corona is above and below the central black hole (Reis \& Miller 2013; Uttley et al. 2014; Chainakun et al. 2019). The corona emits X-ray photons. Observations of some luminous AGNs have shown that X-ray photons contribute a little to the bolometric luminosity. Therefore, as in previous pioneer work, we assume that X-rays contribute 5 percents of the bolometric luminosity (e.g., Liu et al. 2013). The left 95 percents of the bolometric luminosity are mainly optical/UV photons. We further assume that the X-ray photons are spherically emitted. Therefore, the X-ray flux is
\begin{equation}
F_{\rm X}=0.05 L_{\rm bol} \exp{(-\tau_{\rm X})}/4\pi r^2
\end{equation}
$\tau_{\rm X}=\int_0^r \rho \kappa_{\rm es} dr$ is the scattering optical depth, with $\kappa_{\rm es}= 0.4 \ {\rm cm^2 \ g^{-1}}$ being the Thomson scattering opacity.

\begin{figure*}
\begin{center}
\includegraphics[scale=0.5]{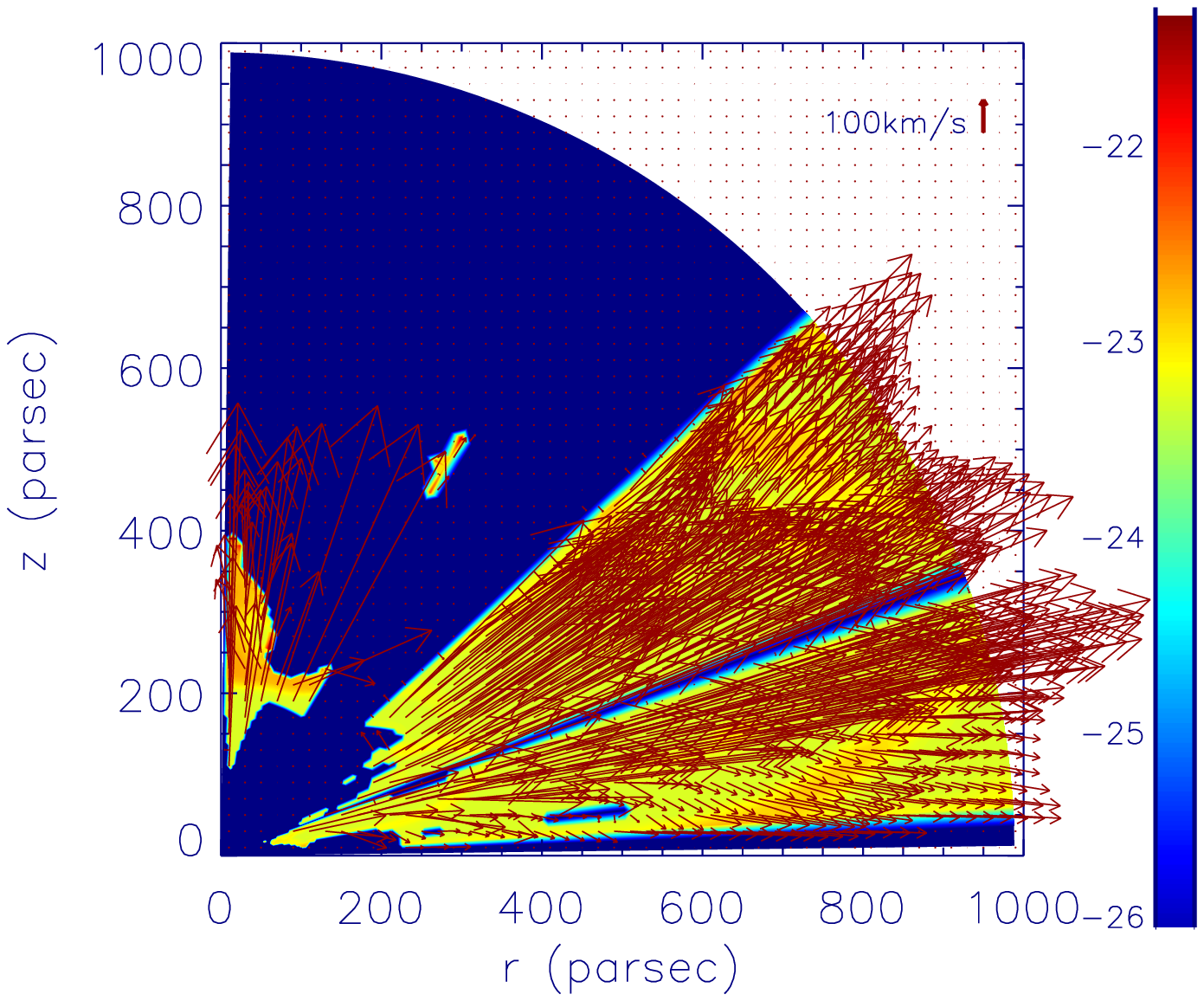}\hspace*{0.1cm}
\includegraphics[scale=0.5]{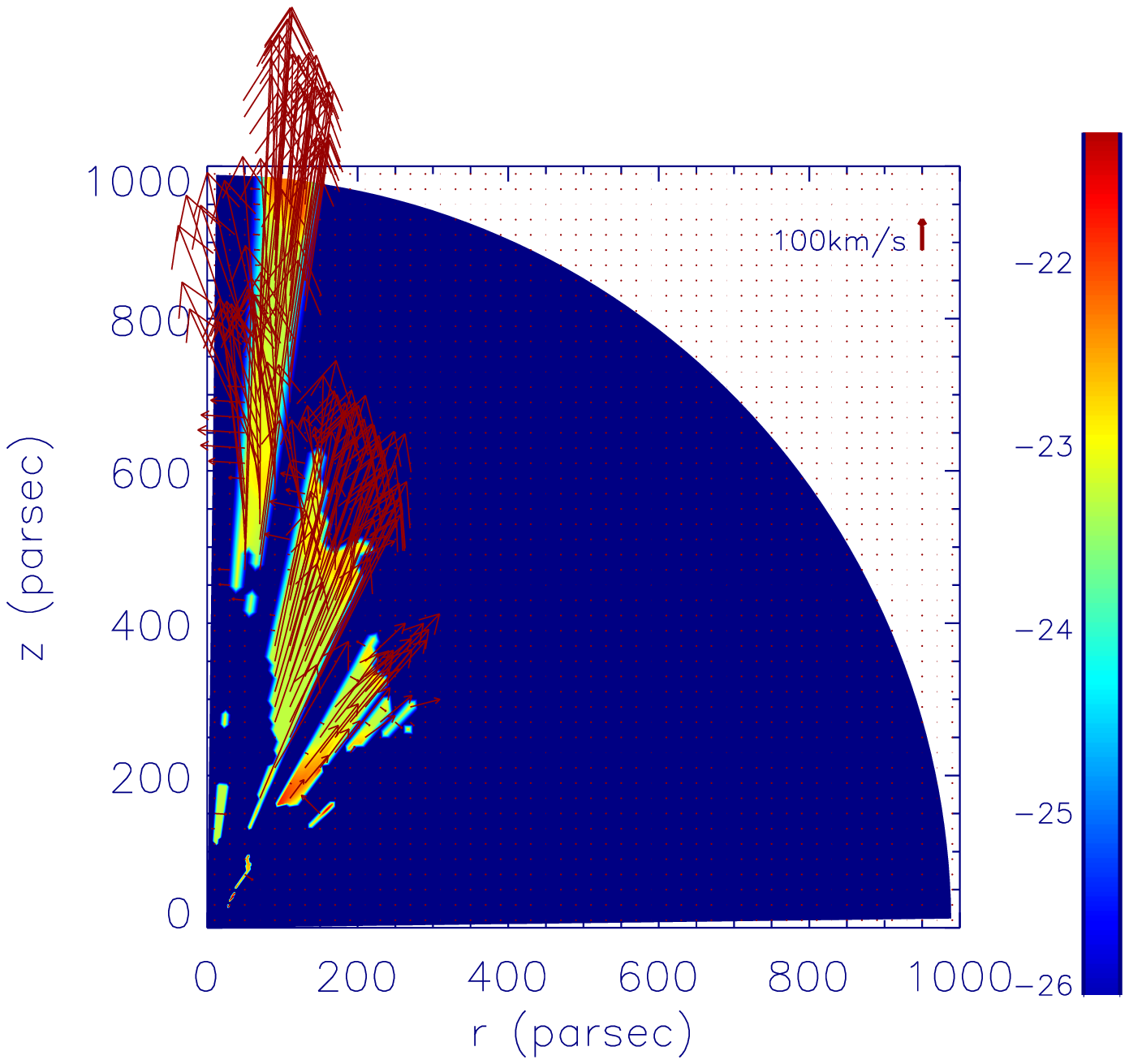}\hspace*{0.7cm}\\
\hspace*{0.5cm} \caption{Distribution of WAs at $5.3 \times 10^7$ year (left panel) and $8.4 \times 10^7$ year (right panel) year for model L0.1v0.3D21T6. Colors denote the logarithm density and arrows show the velocity vector. Note that in this figure, we only show the distribution of WAs. In the regions without WAs, we set the color to dark blue. It does not mean that the regions without WAs have significantly low density. The purpose of plotting this figure in this way is try to protrude the WAs. \label{Fig:vectorWA}}
\end{center}
\end{figure*}

\begin{table*} \caption{Models and results }
\setlength{\tabcolsep}{1.5mm}{
\begin{tabular}{ccccccccc}
\hline \hline
 Model & $\dot m_{\rm inject}$ & $v_{\rm inject}/c$  &  $\rho_0$ & $T_0$ & $C_f$  & $\dot M_{\rm WA}$   & ${\rm \dot E_{kWA}}/L_{\rm bol}$ & ${\rm \dot p}_{\rm WA}/ (L_{\rm bol}/c)$ \\

  & ${L_{\rm Edd}/(0.1 c^2)}$ &  & ($10^{-21}\text{g cm}^{-3}$) & $ 10^5 {\rm K}$ &    & $L_{\rm bol}/(0.1 c^2)$    &  &   \\
(1)    & (2)         & (3)          & (4)         &  (5)  & (6)     &     (7)          &        (8)   & (9)   \\

\hline\noalign{\smallskip}
L0.1v0.1D21T5  & 0.1 & 0.1 & 1 & 1 & $13\%$  &  2.3 & $1.4\times 10^{-4}$ & $6\times 10^{-2}$  \\
L0.1v0.1D22T5  & 0.1 & 0.1& 0.1 & 1 & $6\%$ & 1.2 & $3.6 \times 10^{-5} $ & $3\times 10^{-2}$  \\
L0.1v0.1D23T5  & 0.1 & 0.1& 0.01 & 1 & $10\%$ & 0.7 & $4.5\times 10^{-5} $ & $2\times 10^{-2}$  \\
L0.1v0.1D24T5  & 0.1 & 0.1& 0.001 & 1 & $0.1\%$ & $4\times 10^{-3}$  & $4.8\times 10^{-8}$ & $6\times 10^{-5}$  \\
L0.1v0.1D21T6  & 0.1 & 0.1& 1 & 10 & $11\%$ & 1.4  & $1.1\times 10^{-4}$ & $4\times 10^{-2}$  \\
L0.1v0.1D22T6  & 0.1 & 0.1& 0.1 & 10 & $4\%$ & 0.1  & $6.3\times 10^{-6}$ & $3\times 10^{-3}$  \\
L0.1v0.1D23T6  & 0.1 & 0.1& 0.01 & 10 & $7\%$ & 0.4  & $2.1\times 10^{-5}$ & $1\times 10^{-2}$  \\
L0.1v0.1D24T6  & 0.1 & 0.1& 0.001 & 10 & $6\%$ & 0.4  & $3.3\times 10^{-5}$ & $2\times 10^{-2}$  \\
L0.5v0.1D21T5  & 0.5 & 0.1& 1 & 1 & $6\%$ &  0.3 & $1.9\times 10^{-5}$ & $8\times 10^{-3}$  \\
L0.5v0.1D22T5  & 0.5 & 0.1& 0.1 & 1 & $20\%$ & 4.5 & $4.0\times 10^{-4}$ & $2\times 10^{-1}$  \\
L0.5v0.1D23T5  & 0.5 & 0.1& 0.01 & 1 & $1.7\%$ & $3.7\times 10^{-2}$ & $3.5\times 10^{-6}$ & $1\times 10^{-3}$  \\
L0.5v0.1D24T5  & 0.5 & 0.1& 0.001 & 1 & $0.004\%$ & $2.3\times 10^{-5}$  & $4.3\times 10^{-11}$ & $1\times 10^{-7}$  \\
L0.5v0.1D21T6  & 0.5 & 0.1& 1 & 10 & $47\%$ & 34.9  & $1.1\times 10^{-3}$ & $5\times 10^{-1}$  \\
L0.5v0.1D22T6  & 0.5 & 0.1& 0.1 & 10 & $3\%$ & 1 & $2.0\times 10^{-5}$ & $2\times 10^{-2}$  \\
L0.5v0.1D23T6  & 0.5 & 0.1& 0.01 & 10 & $4\%$ & 0.1 & $1.5\times 10^{-5}$ & $6\times 10^{-3}$  \\
L0.5v0.1D24T6  & 0.5 & 0.1& 0.001 & 10 & $0.005\%$ & $6.8\times 10^{-5}$  & $1\times 10^{-10}$ & $3\times 10^{-7}$  \\
L0.1v0.3D21T5  & 0.1 & 0.3 & 1 & 1 & $23\%$  &  95.6 & $4.4\times 10^{-3}$ & 2.4  \\
L0.1v0.3D22T5  & 0.1 & 0.3& 0.1 & 1 & $52\%$ & 54.5 & $4.7 \times 10^{-3} $ & 2  \\
L0.1v0.3D23T5  & 0.1 & 0.3& 0.01 & 1 & $2\%$ & 0.1 & $2.4\times 10^{-6} $ & $2\times 10^{-3}$  \\
L0.1v0.3D24T5  & 0.1 & 0.3& 0.001 & 1 & $0.01\%$ & $6.4\times 10^{-4}$  & $4.7\times 10^{-8}$ & $2\times 10^{-5}$  \\
L0.1v0.3D21T6  & 0.1 & 0.3& 1 & 10 & $47\%$ & 135  & $3.1\times 10^{-3}$ & 1.8  \\
L0.1v0.3D22T6  & 0.1 & 0.3& 0.1 & 10 & $6\%$ & 0.7  & $7\times 10^{-5}$ & $3\times 10^{-2}$  \\
L0.1v0.3D23T6  & 0.1 & 0.3& 0.01 & 10 & $4\%$ & 0.4  & $7.7\times 10^{-6}$ & $5\times 10^{-3}$  \\
L0.1v0.3D24T6  & 0.1 & 0.3& 0.001 & 10 & $1\%$ & 0.2  & $1.2\times 10^{-5}$ & $6\times 10^{-3}$  \\
L0.5v0.3D21T5  & 0.5 & 0.3& 1 & 1 & $28\%$ &  48 & $3\times 10^{-3}$ & 2  \\
L0.5v0.3D22T5  & 0.5 & 0.3& 0.1 & 1 & $35\%$ & 6.7 & $4.5\times 10^{-4}$ & $2\times 10^{-1}$  \\
L0.5v0.3D23T5  & 0.5 & 0.3& 0.01 & 1 & $0.9\%$ & $2.8\times 10^{-2}$ & $1.8\times 10^{-7}$ & $2\times 10^{-4}$  \\
L0.5v0.3D21T6  & 0.5 & 0.3& 1 & 10 & $29\%$ & 35  & $2\times 10^{-3}$ & 1  \\
L0.5v0.3D22T6  & 0.5 & 0.3& 0.1 & 10 & $4\%$ & 0.6 & $6.9\times 10^{-5}$ & $3\times 10^{-2}$  \\
L0.5v0.3D23T6  & 0.5 & 0.3& 0.01 & 10 & $6\%$ & 0.1 & $1\times 10^{-5}$ & $4\times 10^{-3}$  \\

\hline\noalign{\smallskip}
\end{tabular}}

Note: Col. 1: model names. Col. 2: injected mass flux of UFOs in unit of $L_{\rm Edd}/(0.1c^2)$. Col. 3: velocity of injected UFOs in unit of speed of light. Col 4: the density for initial condition. Col 5. the temperature for initial condition. Col 6. the time-averaged covering factor of WAs. Col 7. the time-averaged mass flux of WAs in unit of the black hole accretion rate. Col. 8: the time-averaged kinetic power of WAs in unit of the bolometric luminosity of the AGN. Col. 9: the time-averaged momentum flux of WAs in unit of $L_{\rm bol}/c$ .
\end{table*}

The last term in Equation (3) is the cooling/heating term. The X-rays from the central black hole region can cool/heat gas by Compton process. The gas can also be heated by photoionization process and cooled by recombination process. Bremsstrahlung cooling and line cooling are also considered. The heating/cooling equations used in this paper are same as those in Proga et al. (2000). The cooling/heating rates are for gas with cosmic abundances. They are as follows,
\begin{equation}
S\rm c=8.9\times10^{-36}n^2(T_X-4T)\xi \ {\rm erg \ cm^{-3} \ s^{-1}}
\end{equation}
\begin{equation}
G\rm x=1.5\times10^{-21}n^2 \xi^{1/4} T^{-1/2} (1-T/T_X) \ {\rm erg \ cm^{-3} \ s^{-1}}
\end{equation}
\begin{equation}
B\rm r=3.3\times 10^{-27}n^2\sqrt{T} \ {\rm erg \ cm^{-3} \ s^{-1}}
\end{equation}
\begin{equation}
\begin{split}
L_{\rm line}=1.7\times10^{-18}{\rm n}^2\exp(-1.3\times10^5/{\rm T})/\xi/\sqrt{\rm T}+\\
            10^{-24} {\rm n}^2 \ {\rm erg \ cm^{-3} \ s^{-1}}
\end{split}
\end{equation}
$S_{\rm c}$, $G_{\rm X}$, $B_{\rm r}$ and $L_{\rm line}$ are Compton heating/cooling, photoionization heating-recombination cooling, bremsstrahlung cooling, and line cooling, respectively. In these equations, ${\rm n}=\rho/\mu m_{\rm p}$ is gas number density, with $\mu=1$ being mean molecular weight and $m_{\rm p}$ being proton mass, respectively. ${\rm T}$ is gas temperature. ${\rm T_X}$ is the Compton temperature of the X-ray photons. As done by Mo\'{s}cibrodzka \& Proga (2013), we set ${\rm T_X}=1.16\times 10^8$ K. $\xi$ is the ionization parameter which is defined as follows,
\begin{equation}
\xi=\frac{4\pi F_X}{\rm n}=0.05 L_{\rm bol} \exp(-\tau_X)/{\rm n} r^2
\end{equation}
The net heating/cooling rate in Equation (3) is expressed as $\rho \dot E=S {\rm c} + G {\rm x} - B {\rm r} - L_{\rm line}$.

\subsection{Initial and boundary conditions}
We study the interaction of UFOs with the ISM in the region from $1 {\rm parsec}$ to $1000 \ {\rm parsec}$. To make the problem as simple as possible, we assume that the ISM has zero angular momentum. Initially, we put ISM with uniform density ($\rho_0$) and temperature ($T_0$) in our computational domain. After a period of evolution, a quasi-steady state of accretion towards the center forms. Then, we inject UFOs from the inner radial boundary. We simulate the region above the midplane, therefore, in $\theta$ direction, we have the domain $0 \leq \theta \leq \pi/2$. Our resolution is $192 \times 64$. In $r$ direction, logarithm grids are adopted in order to well resolve the inner region. In $\theta$ direction, grids are uniformly spaced.

At the inner radial boundary, in the region $\sim 10^\circ - 70^\circ$, we inject UFOs. For other $\theta$ angles, we use outflow boundary conditions. We set the boundary conditions at the outer radial boundary as follows. For a given $\theta$ angle, when we find gas at the last active zone at this angle $v_r < 0 $, then, we inject gas into the computational domain at this angle. The injected gas density and temperature are same as those for the initial conditions. If we find $v_r>0$ at the last active zone at a given $\theta$ angle, then at this angle, outflow boundary conditions are employed. The settings can guarantee continuously gas supplies at the outer boundary. At $\theta=0$, axis-of-symmetry boundary conditions are adopted. At $\theta=\pi/2$, reflecting boundary conditions are employed.

\section{Results}
\begin{figure}
\begin{center}
\includegraphics[scale=0.5]{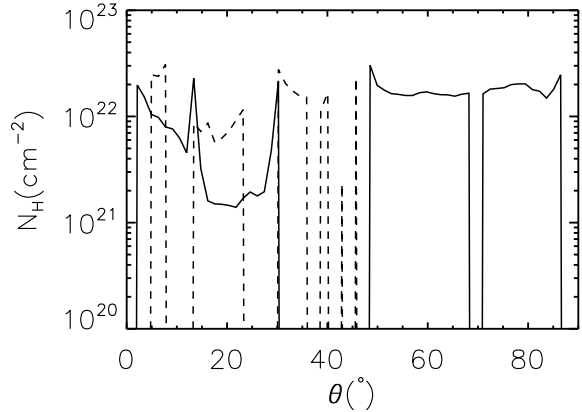}\hspace*{0.7cm}
\hspace*{0.5cm} \caption{Angular distribution of the column density of WAs for model L0.1v0.3D21T6. The solid and dashed lines are for $5.3\times 10^7$ year and $8.4\times 10^7$ year, respectively. \label{Fig:NH-WA}}
\end{center}
\end{figure}

We initially set uniform density and temperature in our computational domain. We first run the simulation without UFOs injection. We find that after a period of $\sim 3.8\times 10^7 {\rm year}$, the flow will achieve a quasi-steady state. The indication of the quasi-steady state is that the mass accretion rate measured at the inner radial boundary does not vary much with time. At the quasi-steady state, we inject UFOs at the inner boundary and mark the time when we inject UFOs as ${\rm time} = 0 {\ \rm year}$. After another period, we will find another quasi-steady state. We judge the quasi-steady state as follows. We calculate the total mass outflow rate at the outer radial boundary. When we find that the mass outflow rate does not monotonously decrease or increase with time, but oscillates around a mean value, we think a quasi-steady state has been achieved. We find that after another period of $\sim 3.8\times 10^7 {\rm year}$, the flow will set down to another quasi-steady state. We mark the time at the beginning of the new quasi-steady state as ${\rm time} = 3.8\times 10^7 {\ \rm year}$. We time-average the properties of WAs from ${\rm time} = 3.8\times 10^7 {\ \rm year}$ to ${\rm time} = 9.8\times 10^7 {\ \rm year}$.

We recognize the WAs features to be detected if the following conditions are satisfied: (A) the velocity of outflowing gas with $-1 \leq \log \xi \leq 3 \ {\rm erg\ s^{-1}}$ is in the range $100-2000 {\rm km \ s^{-1}}$. (B) The column density of the outflowing gas satisfying condition (A) is in the range $N_{\rm H} \sim 10^{20}-10^{22.5} {\rm cm^{-2}}$.

\begin{figure}
\begin{center}
\includegraphics[scale=0.5]{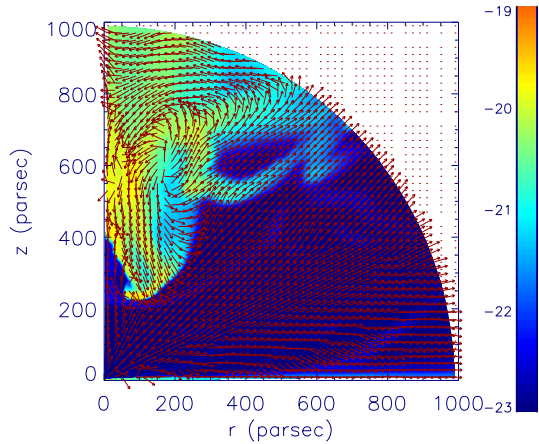}\hspace*{0.7cm}
\hspace*{0.5cm} \caption{Two-dimensional properties of the flow at $5.3 \times 10^7$ year for model L0.1v0.3D21T6. Colors denote the logarithm density (color) and arrows show the unit velocity vector. \label{Fig:vector}}
\end{center}
\end{figure}

The column density of WAs ($N_{\rm H} (\theta)$) is a function of viewing angles between $\theta = 0^\circ$ and $90^\circ$. The covering factor of WAs are calculated as follows,
\begin{equation}
C_f = 4\pi \int^{\pi/2}_0 C_{\rm WA}(\theta) \frac{\sin{\theta} d\theta}{4\pi}
\end{equation}
The parameter $C_{\rm WA}(\theta)$ is defined as follows. At a given $\theta$ angle, if we find that $N_{\rm H} (\theta) = 0$, then we set $C_{\rm WA}(\theta)=0$; when $N_{\rm H} (\theta) \neq 0$, we set $C_{\rm WA}(\theta)=1$. The parameter $C_{\rm WA}(\theta)$ guarantees that we only integrate over viewing angles that having WAs.

The mass flux, kinetic power, and momentum flux of WAs are calculated as follows,
\begin{equation}
\dot M_{\rm WA} = 4\pi r^2 \int^{\pi/2}_0 C_{\rm WA}(\theta) f_{\rm WA} \rho \max(v_r, 0) \sin{\theta} d\theta
\end{equation}
\begin{equation}
\dot E_{\rm kWA} = 2\pi r^2 \int^{\pi/2}_0 C_{\rm WA}(\theta) f_{\rm WA} \rho \max(v_r^3, 0) \sin{\theta} d\theta
\end{equation}
\begin{equation}
\dot {\rm p_{\rm WA}} = 4\pi r^2 \int^{\pi/2}_0 C_{\rm WA}(\theta) f_{\rm WA} \rho \max(v_r^2, 0) \sin{\theta} d\theta
\end{equation}
In Equations (11)-(13), the parameter $f_{\rm WA}$ is defined as follows. If the outflowing gas at a given computational grid has $-1 \leq \log \xi \leq 3 \ {\rm erg\ s^{-1}}$ and $100 \leq v \leq 2000 {\rm km \ s^{-1}}$, we set $f_{\rm WA}=1$. Otherwise, we set $f_{\rm WA}=0$. This can guarantee that the quantities calculated by Equations (11)-(13) are only for WAs.

As shown below, in addition to WAs, UFOs can also drive non-WAs outflows. The properties of the non-WAs outflowing gas (column density, ionization parameter and velocity) do not satisfy the requied conditions of WAs. We define the mass flux, kinetic power, momentum flux of all the outflows (including WAs and non-WAs) as $\dot M_{\rm total}$, $\dot E_{\rm total}$ and $\dot p_{\rm total}$, respectively. We can obtain the values of $\dot M_{\rm total}$, $\dot E_{\rm total}$ and $\dot p_{\rm total}$ by just directly using Equations (11)-(13) with the dropping of the factor of $C_{\rm WA}(\theta) \ f_{\rm WA}$.

There are two ways to obtain time-averaged values of WAs. We take the mass flux of WAs as an example. The first way is as follows. we first time-average the density and velocity. Then, we use the time-averaged density and velocity to calculate the mass flux of WAs. In the second way, we first get the mass flux of the WAs at each snapshot. Then, we time-average the mass flux. If the flow is much laminar, the two ways will give similar results. However, if the flows are quite turbulent, the two ways will give totally different results. For a turbulent system, the first way will erase important information about the true nature of the flow. In this paper, we find that the flow is quite turbulent, therefore, we choose the second way to obtain the time-averaged values of the WAs in Table 1.

We summarize all the models in Table 1. Column 2 gives the injected mass flux of UFOs in unit of $L_{\rm Edd}/(0.1c^2)$. Column 3 gives the velocity of injected UFOs in unit of speed of light. Columns 4 and 5 give the initial density and temperature, respectively. Column 6 gives the time-averaged covering factor of WAs. Column 7 gives the time-averaged mass flux of WAs in unit of the black hole accretion rate. Column 8 gives the time-averaged kinetic power of WAs in unit of the bolometric luminosity of the AGN. Column 9 gives the time-averaged momentum flux of WAs in unit of $L_{\rm bol}/c$.

From Table 1, we can see that generally models with higher density have stronger WAs. We take model L0.1v0.3D21T6 which has strong WAs as our fiducial model to study the properties of WAs.

\subsection{Model ${\rm L0.1v0.3D21T6}$}
We find that the angular location of WAs is changing with time. In order to illustrate this result, we plot Figure \ref{Fig:vectorWA}. This figure shows the distributions of WAs. Colors are for logarithm density of WAs. Arrows show the velocity of WAs. In this figure, the left and right panels are for $5.3 \times 10^7$ year and $8.4 \times 10^7$ year, respectively. Note that in this figure, we only show the distribution of WAs. In the regions without WAs, we set the color to dark blue. It does not mean that the regions without WAs have significantly low density. The purpose of plotting this figure in this way is try to protrude the WAs. It can be easily seen that at $5.3 \times 10^7$ year, WAs are mainly distributed in the region $48^\circ < \theta < 82^\circ$. There is also little amount of WAs in the region $\theta < 30^\circ$. At $8.4 \times 10^7$ year, WAs are distributed in the region $\theta < 45^\circ$. Figure \ref{Fig:NH-WA} shows the angular distribution of the column density of WAs at $5.3 \times 10^7$ year (solid line) and $8.4 \times 10^7$ year (dashed line). This figure also indicates the changing of angular locations of WAs with time. For an AGN, our viewing angle is fixed. Our results indicate that for an AGN, WAs can be observed intermittently. For the observations do not show signiture of WAs, it might not mean that WAs are not present. Instead, WAs may move out of our line of sight.

There are two possible reasons for the changing of angular distributions of WAs. The first reason may be that the angular locations of WAs do vary with time. In order to illustrate the second reason, we plot Figure \ref{Fig:vector}. In this figure, we plot the two-dimensional properties of the flow at $5.3 \times 10^7$ year. Colors show logarithm density including both inflows and outflows. Arrows show unit velocity vector. Comparing this figure to the left panel of Figure \ref{Fig:vectorWA}, we can see that WAs are just parts of the total outflows. For example, there are outflows at $\sim 38^\circ$ close to the outer radial boundary. The outflows at this location do not satisfy the conditions of WAs. We call the outflows do not satisfy the conditions of WAs as non-WAs outflow. If the outflows here keep present but with the properties changing with time, the non-WAs outflows may become WAs. Therefore, the second possible reason for the changing of angular distributions of WAs is that due to the changing of properties, the WAs/non-WAs outflow can become non-WAs/WAs outflows.

\begin{figure*}
\begin{center}
\includegraphics[scale=0.7]{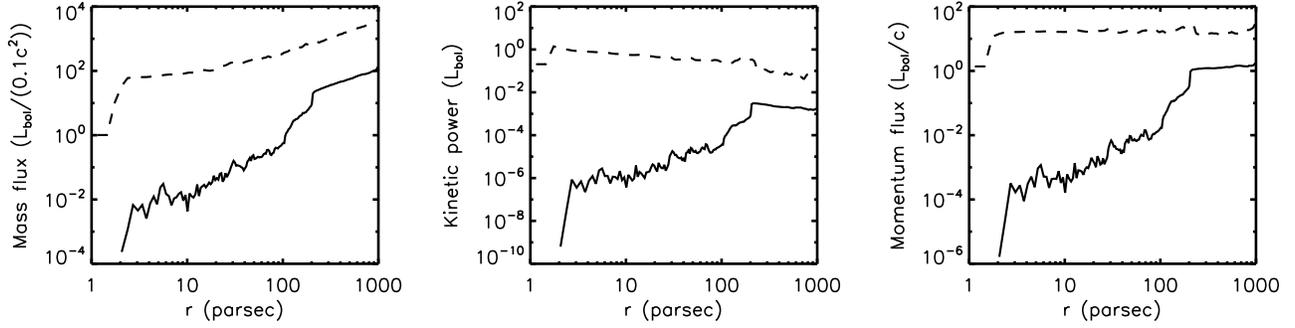}\hspace*{0.7cm}
\hspace*{0.5cm} \caption{Time-averaged radial profiles of mass flux (left panel), kinetic power (middle panel) and momentum flux (right panel) of outflows. The solid line is for WAs. The dashed line is for total outflows. \label{Fig:flux}}
\end{center}
\end{figure*}

The time-averaged covering factors of the WAs listed in Table 1 are calculated as follows. We first calculate the covering factor at each snapshot. Then, we do the time-average for the covering factors at different snapshots. The time-averaged covering factor is $47\%$.

A interesting question is that how much mass can be carried away by WAs. We plot the radial profile of the mass flux of WAs in the left panel (solid line) of Figure \ref{Fig:flux}. The mass flux of WAs increases with the increase of radius. This means that when WAs move outwards, more and more gas becomes WAs. The mass flux of WAs is a function of radius. The mass flux of WAs listed in Table 1 is its maximum value. The maximum mass flux of WAs can be more than two orders of magnitude higher than the black hole accretion rate. As introduced above, in addition to WAs, there are also non-WAs outflows. In the left panel of Figure \ref{Fig:flux}, we also plot the mass flux of all the outflows (including both WAs and non-WAs) using a dashed line. The mass flux of all the outflows is more than one order of magnitude higher than that of the WAs.

\begin{figure*}
\begin{center}
\includegraphics[scale=0.5]{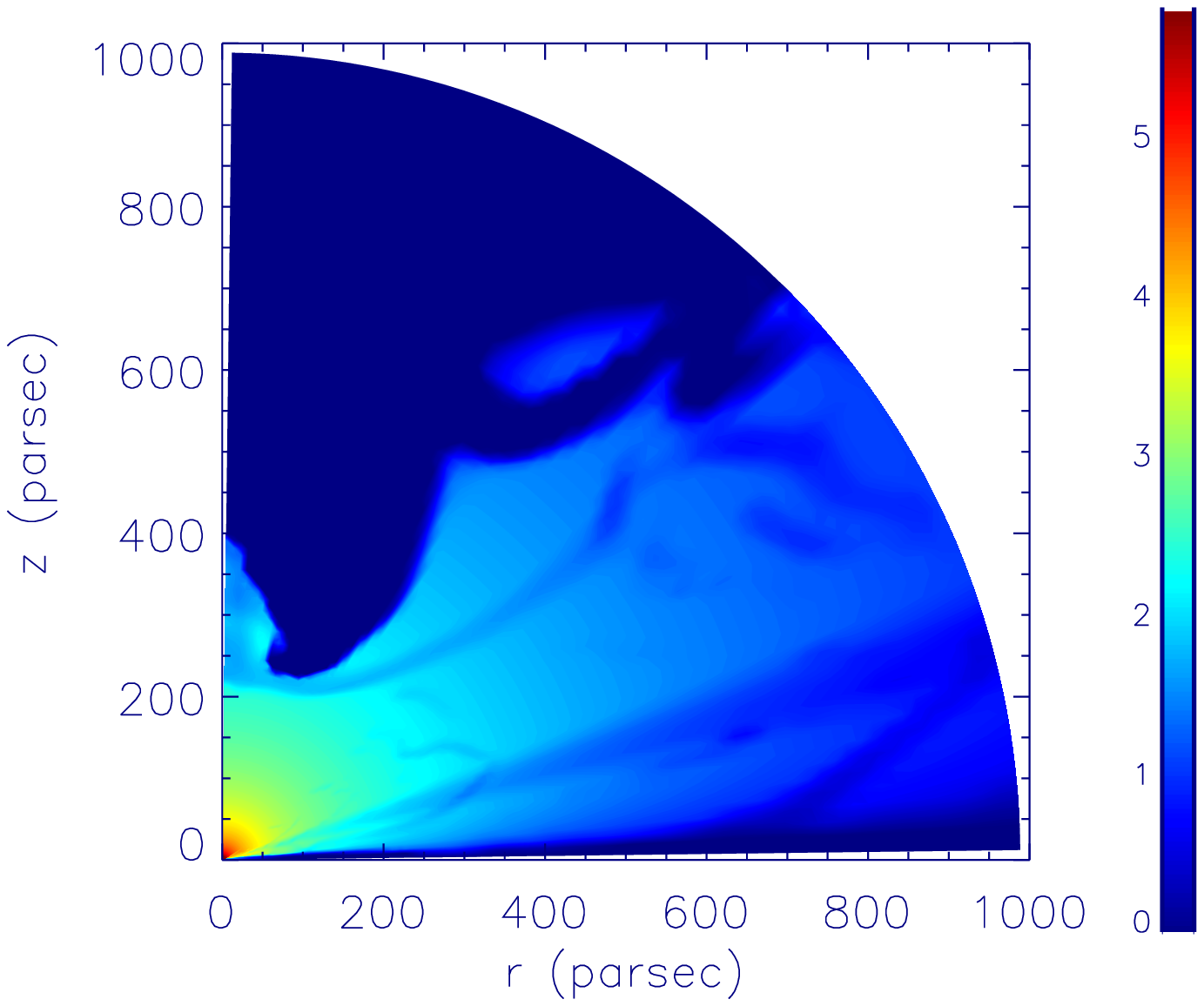}\hspace*{0.1cm}
\includegraphics[scale=0.5]{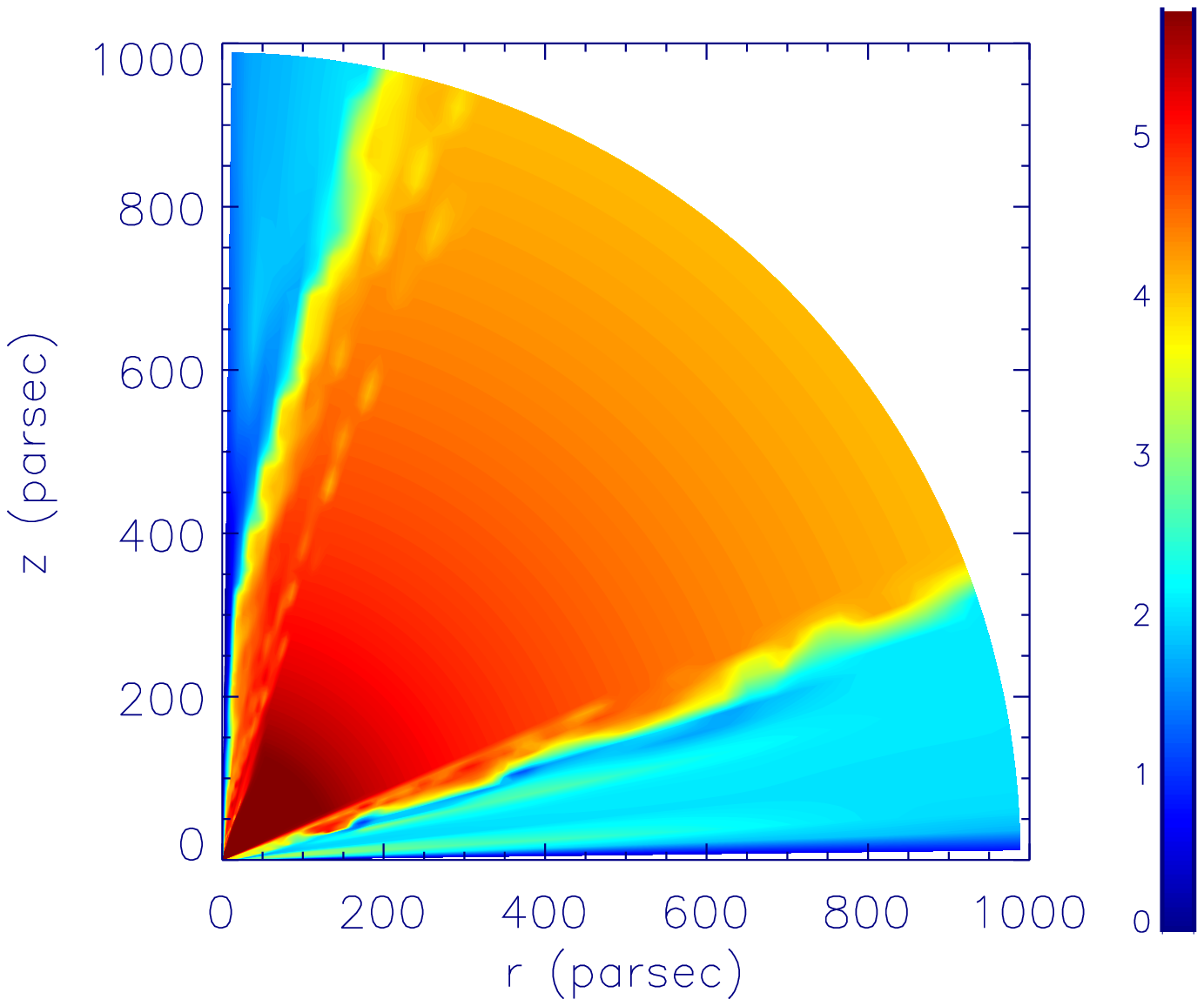}\hspace*{0.7cm}\\
\hspace*{0.5cm} \caption{Two-dimensional distributions of logarithm of ionization parameter of gas at $5.3 \times 10^7$ year for model L0.1v0.3D21T6 (left panel) and L0.1v0.3D24T6 (right panel). \label{Fig:xi}}
\end{center}
\end{figure*}

The solid line in the middle panel of Figure \ref{Fig:flux} denotes the kinetic power of WAs. The kinetic power of WAs is lower than $0.5 \%$ of the bolometric luminosity. It has been shown that an outflow with kinetic power larger than $0.5\% L_{\rm bol}$ can give sufficient feedback to its host galaxy (e.g., Di Matteo et al. 2005; Hopkins \& Elvis 2010). Therefore, the WAs can not play some important role in regulating the properties of its host galaxy. The kinetic power of the UFOs injected at the inner radial boundary is $1/2 \dot m_{\rm inject} v_{\rm inject}^2 \sim 0.5 L_{\rm bol}$. The kinetic power of all of the outflows ($\dot E_{\rm total}$) is showed by using the dashed line in the middle panel of Figure \ref{Fig:flux}. It can been seen that inside $10 \ \rm parsec$, $\dot E_{\rm total}$ is comparable to that of the injected UFOs. Also, we note that inside $10 \ \rm parsec$, the mass flux of total outflow $\dot M_{\rm total}$ is significantly larger than that of UFOs. The results indicate that the kinetic energy of the UFOs is very effectively transferred into the ISM.  $\dot E_{\rm total}$ decreases with increase of radius. This is because that when moving outwards, the gravity of the black hole do negative work to the outflows. Parts of the kinetic energy of outflows are transferred into their gravitational energy. At the radial outer boundary, $\dot E_{\rm total} \sim 10\% L_{\rm bol}$. Therefore, the UFO driven outflows can have a significant influence on the evolution of its host galaxy.

The kinetic power of outflow is $\dot E = 1/2\dot M v^2$. The momentum flux of outflow is $\dot p = \dot M v$. Therefore, we have $\dot p = 2\dot E/v $ and $\dot p/(L_{\rm bol}/c) = \frac{2 \dot E}{L_{\rm bol}} \frac{c}{v}$. From the middle panel of Figure \ref{Fig:flux}, we can see that $\dot E_{\rm total} \sim L_{\rm bol}$. Also, the velocity of the outflows is significantly smaller than the speed of light. Therefore, the momentum flux of the outflow $\dot p_{\rm total}$ can be significantly higher than $L_{\rm bol}/c$ (see the right panel of Figure \ref{Fig:flux}). The UFOs driven outflow can go to larger radii to interact with the gas in its host galaxy. If the host galaxy gas can very quickly cool down after interacting with the outflows and photons from the central AGN, the behaviour of the gas will be controlled by the momentum carried by the radiation and outflow from the AGN. In such case, the outflows should play much more important role in regulating the properties of the galaxy, because the momentum flux of outflows is significantly higher than that of the radiation.

\begin{figure}
\begin{center}
\includegraphics[scale=0.5]{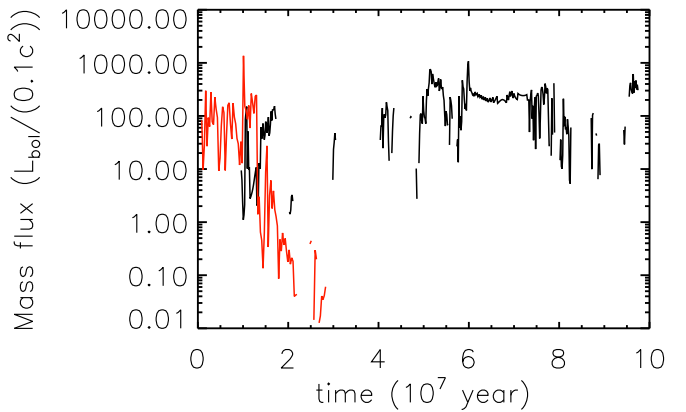}\hspace*{0.7cm}
\hspace*{0.5cm} \caption{Time evolution of the mass flux of WAs measured at the outer radial boundary. The black and red lines are for models L0.1v0.3D21T6 and L0.1v0.3D21T6-Bn, respectively. \label{Fig:WA-compare}}
\end{center}
\end{figure}
\subsection{Effects of changing properties of gas}
We now investigate the effects of changing the properties of gas. The results in Table 1 show that generally, with the decrease of the gas density, the strength (mass flux, kinetic power, momentum flux) of WAs decreases. The reason is that with the decrease of gas density, the ionization parameter will become larger. The gas is more easily to have ionization parameter larger than the upper limit of the ionization parameter of WAs ($\log \xi = 3$). We take models L0.1v0.3D21T6 and L0.1v0.3D24T6 to do comparisons. Figure \ref{Fig:xi} plots the two-dimensional distribution of $\log \xi$ at $5.3 \times 10^7$ year for model L0.1v0.3D21T6 (left panel) and L0.1v0.3D24T6 (right panel). It is clear that the gas in model L0.1v0.3D24T6 has higher ionization parameter. In a significant larger region in model L0.1v0.3D24T6, the gas has ionization parameter much higher than the upper limit of the ionization parameter of WAs ($\log \xi = 3$). Therefore, models with lower gas density tend to generate weaker WAs.

From Table 1, we see that the initial gas temperature has very little effects on the strength of WAs. The reason is as follows. The mass flux of WAs increases with the increase of radius (see Figure \ref{Fig:flux}). Therefore, most of the WAs are generated at large radii. At large radii ($> 100$ parsec), we find that the cooling/heating timescale calculated as $e/\rho \dot E$, is significantly smaller than the escape timescale ($r/v_r$) of WAs. Therefore, the internal energy of gas can be quickly changed before gas can move a large distance. In other words, because of the quick cooling/heating, gas can quickly lose the memory of its initial internal energy. Thus, the initial gas temperature has little effects on the strength of WAs.

\subsection{Effects of strength of UFOs}
We discuss the effects of strength of UFOs on the properties of WAs. We can change both the velocity and mass flux of UFOs injected at the inner radial boundary. From Table 1, it is clear that generally when the initial gas density is high ($10^{-21} {\rm g \ cm^{-3}}$), with the increase of the velocity or mass flux of UFOs, the WAs will become stronger. This is easy to understand that stronger UFOs can drive stronger WAs. However, we note that the increase of strength of WAs with the increase of strength of UFOs is not significant. When the initial gas density is lower than $10^{-21} {\rm g \ cm^{-3}}$, the effects of changing the strength of UFOs are extremely weak. The reason is as follows. When the initial gas density is much lower, as introduced above, the ionization parameter of gas will be much higher. In this case, the ionization parameter of gas is too higher that only a tiny fraction of the gas can satisfy the ionization parameter criterion of WAs. For such high ionization parameter gas, the increase of strength of UFOs can hardly increase the strength of WAs. The increase of strength of UFOs does help drive stronger outflows. However, most fraction of the stronger outflows still not satisfy the ionization criterion of WAs.

\subsection{Effects of outer boundary conditions}
In this paper, in order to achieve a quasi-steady state, we allow gas to freely move out/into the computational domain at the outer radial boundary (see Section 2.3). This kind of settings can keep continuous supplying of gas into the computational domain from the outer radial boundary. We do find that quasi-steady state can be achieved by using such kind of boundary conditions. The time evolution of the mass flux of the WAs measured at the radial outer boundary for our fiducial model L0.1v0.3D21T6 is shown by the black line in Figure \ref{Fig:WA-compare}. It is clear that the mass flux oscillates around some mean values. A quasi-steady state has been achieved in this model. Note that at some snapshots, the mass flux of WAs at the outer boundary is 0. This is because WAs are not smoothly distributed in the computational domain. They are much clumpy (see Figure \ref{Fig:vectorWA}). There are no WAs at the outer boundary for these snapshots. Another kind of usually adopted boundary condition is the outflow boundary condition. Outflow boundary condition allows gas to freely flow out of the computational domain, but gas is not permitted to go into the computational domain from the boundaries. Therefore, the total mass in the computational domain will keep decreasing with time. We carry out a model L0.1v0.3D21T6-Bn to test the effects of such boundary conditions. The only difference between model L0.1v0.3D21T6-Bn and our fiducial model L0.1v0.3D21T6 is the setting of outer radial boundary conditions. In model L0.1v0.3D21T6-Bn, at the outer radial boundary, we use outflow boundary conditions to not allow gas to flow into our computational domain. The mass flux of WAs measured at the outer radial boundary is shown by the rad line in Figure \ref{Fig:WA-compare}. It is clear that the mass flux decreases with time. After $3\times 10^7$ year, the WAs disappear. The reason is as follows. The UFOs blow out the overlying ISM continuously, without newly injected gas into the computational domain, the only left gas is the UFOs injected from the inner radial boundary. From the analysis above, we conclude that the properties of the WAs mainly depend on the density of the ISM injected into the computational domain from the outer radial boundary.

In a real galaxy, there may be two cases for gas supplies from the galaxy scale to the central AGN. In the first case, gas in the galaxy scale continuously supplies to the central AGN. This case is mimicked by the free inflow/outflow outer boundary conditions used in the models in Table 1. In the second case, there is no gas supplies from the galaxy scale to the central AGN. This case is mimicked by the boundary conditions used in model L0.1v0.3D21T6-Bn.

In our models, we assume that UFOs are always present. In reality, the UFOs may be intermittently produced by the AGN. For the intermittently generated UFOs, there are two related timescales. The first one is how long the UFOs can last once it is produced. The second one is that after the UFOs disappearing, how long it will take for another UFOs to be generated. There is no information about the two timescales from both observations and theory.

We discuss the possible effects of intermittent UFOs in the case that there is no continuously gas supplies from the galaxy scale to the central AGN (boundary conditions in model L0.1v0.3D21T6-Bn). If the UFOs do not last sufficiently long time to push the gas away from our computational domain, the gas with velocity smaller than the escape velocity may fall back after UFOs's shutting down. If the shutting down timescale of UFOs is not so long, the fall back gas may not have sufficient time to go into the region inside 1 parsec. If both the two conditions are satisfied, the WAs may be generated periodically. On a time-averaged sense, this will produce more WAs compared to continuously UFO injection (model L0.1v0.3D21T6-Bn). However, it is not clear whether the two conditions can be satisfied simultaneously. Therefore, it is hard to judge whether can intermittent UFOs produce more WAs.

In the case that there is continuously gas supplies from galaxy scale to the central AGN (the models in Table 1), due to the lack of information about the duration timescale of UFOs, it is also quite hard to guess the effects of intermittent UFOs on the WAs.

\subsection{Comparison to observations}
\subsubsection{covering factor and launching mechanisms}
WAs are detected in $65\%$ of the nearby AGNs (Laha et al. 2016). Assuming that the nearby AGNs are observed with random viewing angles, the covering factor of WAs in these sources is $65\%$. In this paper, we find that the covering factor of WAs varies with numerical settings of the simulations. The covering factor of WAs mainly depends on the gas density. For models with high gas density, the covering factor of WAs can reach a value as high as $52\%$ (model L0.1v0.3D22T5). For models with much lower gas density, the covering factor of WAs can be lower than $0.01\%$. The gas density in the nearby AGNs which show the existence of WAs should vary significantly from source to source. Therefore, in these sources, UFOs may play some role in driving WAs. However, this scenario is not sufficient to explain the high covering factor of WAs in these sources. Other WA driving mechanisms should be at work in these sources.

One possible mechanism is the magneto-centrifugal outflow model (Blandford \& Payne 1982). This model requires that a large scale open magnetic field should be present. The magnetic field is anchored in a thin accretion disk at the midplane. Also, it is required that the angle between the magnetic field line and the accretion disk should be smaller than $60^\circ$. An accretion disk can be present inside hundreds or thousands of Schwarzschild radii. At larger radii, the accretion disk should be gravitationally unstable. Observations show that the distance of WAs to the AGN center has a very broad distribution. The minimum location of WAs can be as small as 0.1 parsec. The maximum location of WAs can reach 1000 parsec or beyond (Laha et al. 2016). We note that, the observed location of WAs should be the upper limit of the distance to the center where the WAs are launched. In other words, it is possible that WAs are launched at a smaller distance to the black hole. After be launched, WAs move to larger radii and then be detected by observations. If the WAs are launched at the accretion disk scale, then the MHD scenario may be applicable. In this case, at the launching point, the outflow may be not recognized as WAs. For example, due to the close distance to the center of the AGN, its ionization parameter may be too high to satisfy the conditions of WAs. When the outflows move to large radii, the properties of outflows (ionization parameter, velocity) may change into the scope of WAs. There is another possibility that WAs are launched at the location where it is observed. In this case, there should be no stably existing accretion disk at the midplane. Consequently, it should be hard to keep an appropriate angle between magnetic field lines and the disk to launch outflows. In this case, the magneto-centrifugal model should be problematic.

The second possible mechanism is the thermally driven outflow model. Mizumoto et al. (2019) studied this mechanism. It is found that the thermal driven outflow model can successfully predict the velocity of the WAs. However, the column density of the thermally driven outflows is lower than the observed value. It is suggested that additional outflow launching mechanism is needed for the WAs (Mizumoto et al. 2019).

The third mechanism is the radiation pressure driven outflow model. The nearby AGNs are mainly sub-Eddington sources. Therefore, the radiation pressure due to Thomson scattering is not possible to driven WAs. In addition, the warm absorbers have temperature around $10^6$ K. The radiation pressure due to spectral lines (line force) can not play a role for such a temperature (Proga 2007). The opacity of dust is much larger than that of the electron scattering, so there is a `force multiplier' effect, which reduces the effective Eddington limit. Fabian et al. (2008) find that the cross-section of dust can be $\leq 100$ for column density $\sim < 10^{21} {\rm cm^{-2}}$. Therefore, the bright nearby AGNs are effectively super-Eddington for low column density dusty gas. Radiation pressure on dust may play some role in driving WAs. However, the WAs with column density $> 10^{21} {\rm cm^{-2}}$ can not be driven by radiation pressure on dust.

In summary, it seems that the WAs can not be driven by a solely mechanism. All the four mechanisms (UFOs driven, magneto-centrifugal driven, thermal driven, radiation pressure driven) may play some role in driving WAs.

\subsubsection{mass flux and kinetic power}
It is quite hard to directly give the mass flux of WAs by observations. The formula often used to estimate the mass flux of outflows in observational literatures is as follows,
\begin{equation}
\dot M \sim \mu m_{\rm p} {\rm n} r^2 v(r) \Omega
\end{equation}
$\Omega$ is the opening solid angle of the WAs. The value of $\Omega = C_f \ 4\pi$. Further assuming that ${\rm n} r \sim N_{\rm H}$, the above equation can be re-written as,
\begin{equation}
\dot M \sim 4 \pi C_f \mu m_{\rm p} r N_{\rm H} v(r) C_v
\end{equation}
$C_v$ is the volume filling factor of WAs. For an AGN, the viewing angle is fixed. One can only obtain the column density along the line of sight. The column density along other directions is not known. Therefore, one should use the parameter $C_v$ to estimate the mean value of the column density. The mean column density equals $C_v N_{\rm H}$. The velocity of WAs can be obtained by using the information of the absorption lines. There are some methods to estimate the location of the absorbers (see Blustin et al. 2005). The only unknown parameter in the above equation is $C_v$. The value of $C_v$ is quite hard to estimate. It must be less than or equal to 1. Blustin et al. (2005) analyzed the properties of WAs in a sample of Seyfert type AGNs. By assuming that the WAs are launched by radiation pressure and setting the momentum flux of the WAs equaling to the radiation momentum, Blustin et al. (2005) found that the value of $C_v$ is
never more than $8\%$. It is found that in $70\%$ of the AGNs in their sample, the mass flux of outflow is greater than the black hole accretion rate. In most of the AGNs in their sample, the kinetic power ($1/2\dot M v^2$) of WAs is found to be significantly less than $1\%$ of the bolometric luminosity.

In the present paper, we find that the mass flux of WAs in some sources can be higher than the black hole accretion rate. However, the kinetic power of WAs in all the models is lower than $1\%$ of the bolometric luminosity (see Table 1). The consistence of our simulation results with the observation results (Blustin et al. 2005) may be a coincidence. The consistence suggests that WAs either by radiation pressure driven (Blustin et al. 2005) or UFOs driven can not supply sufficient feedback to its host galaxy.

\section{Discussion}
The optical depth is used to calculate the ionization parameter of the outflows. When we calculate the optical depth, we only consider the contribution of gas outside 1 parsec. Actually, when the radiation from the vicinity of the central black hole propagates outwards, some portion of photons should be scattered into other directions by the gas inside 1 parsec. It is generally believed that inside 1 parsec, at the midplane, there is a thin disk; above and below the midplane, there is outflow. For the direction along the midplane, the optical depth inside 1 parsec is contributed by the thin disk. It should be much larger than 1. In this case, for the direction along midplane, the optical depth may be significantly larger than that used in this paper. However, the thin disk inside 1 parsec should be significantly thin. Therefore, it can only affect the calculation of optical depth in a negligibly small angular region.  Above and below the midplane, if there are just UFOs outflows, the optical depth contributed by UFOs above and below the midplane is $\sim \mu m_{\rm p} \rho dr \sim \mu m_{\rm p} N_{\rm H}$. Given that the column density of UFOs $<10^{24} \ {\rm cm^{-2}}$ (Tombesi et al. 2011), the optical depth contributed by the gas inside 1 parsec should be smaller than 1. In this case, the contribution to optical depth in the region above and below the midplane by the gas inside 1 parsec is not important.  In addition to UFOs, if there are other types of outflows below and above the midplane inside 1 parsec, the optical depth used in this paper is further underestimated. However, because the ionization parameter of WAs spans four orders of magnitude, the non-accurate optical depth used in this paper should not significantly affect the recognition of WAs.

The gas is assumed to have zero angular momentum. Therefore, the centrifugal force of gas is neglected. The strength of WAs may be underestimated. We note that the omission of centrifugal force may not significantly affect the acceleration of outflows. The reason is as follows. If we assume angular momentum of outflows is conserved when outflows move outwards, we have $v_{\phi} \propto r^{-1}$. The centrifugal force $v_{\phi}^2/r \propto r^{-3}$. The gravitational force $\propto r^{-2}$. The maximum rotational velocity of gas equals to the Keplerian value. Even if gas has Keplerican angular momentum, the centrifugal force only contributes significantly at the starting location of the outflows. When outflows move outwards, the centrifugal force decreases faster than gravity. Therefore, the omission of centrifugal force may not severely affect our results.

\section{Summary}
We perform two-dimensional hydrodynamic simulations to study whether the WAs found in nearby AGNs can be driven by the UFOs. We find that when UFOs collide with the ambient gas, WAs can be generated. The properties (covering factor, mass flux, kinetic power) of WAs mainly depend on the density of the ISM injected into the computational domain from the outer radial boundary. Higher density gas can generate higher covering factor WAs. The highest covering factor of WAs found in this paper is $\sim 50\%$ when the ambient gas has a initial density $10^{-21} {\rm g \ cm^{-3}}$. When the gas has much lower density, the covering factor will become smaller. Observations show that the WAs in nearby AGNs have covering factor $65\%$ (Laha et al. 2016). Therefore, we conclude that UFOs should play some role in driving WAs. However, this scenario alone is not sufficient to explain observations. Other mechanisms should also contribute to the formation of WAs. We also find that the WAs driven by UFOs have kinetic power much lower than $1\%$ of the bolometric luminosity of its host AGN. Therefore, the UFOs driven WAs might not affect the evolution of its host galaxy (e.g., Di Matteo et al. 2005; Hopkins \& Elvis 2010).

\section*{Acknowledgments}
D.-F. Bu is supported by the Natural Science Foundation of China (grant 11773053). X.-H. Yang is supported by the Natural Science Foundation of China (grant 11973018) and the Chongqing Natural Science Foundation (grant cstc2019jcyj-msxmX0581). This work made use of the High Performance Computing Resource in the Core Facility for Advanced Research Computing at Shanghai Astronomical Observatory.

\end{document}